\documentclass[sigconf]{acmart}

\renewcommand\footnotetextcopyrightpermission[1]{}
\setcopyright{none}
\makeatletter
\def\@conference{}

\makeatother

\usepackage{amsmath}
\usepackage{newtxmath}
\usepackage{amsthm}

\usepackage{graphicx}
\usepackage{booktabs}
\usepackage{multirow}
\usepackage{array}
\usepackage{tabularx}
\newcolumntype{Y}{>{\RaggedRight\arraybackslash}X}

\usepackage[ruled,vlined,linesnumbered]{algorithm2e}

\usepackage{xcolor}
\usepackage{listings}

\lstdefinelanguage{Solidity}{
    keywords={contract, function, public, payable, mapping, address, uint256, msg, sender},
    keywordstyle=\color{cyan!80!black}\bfseries,
    morestring=[b]",
    stringstyle=\color{orange!80!black},
    morecomment=[l]{//},
    morecomment=[s]{/*}{*/},
    commentstyle=\color{gray!60}\itshape,
    sensitive=true
}

\lstdefinestyle{code-common}{
    basicstyle=\ttfamily\small\color{black},
    keywordstyle=\color{cyan!80!black}\bfseries,
    commentstyle=\color{gray!60}\itshape,
    stringstyle=\color{orange!80!black},
    frame=single,
    rulecolor=\color{black},
    backgroundcolor=\color{white},
    breaklines=true,
    tabsize=4,
    showstringspaces=false,
    numbers=left,
    numberstyle=\tiny\color{gray!70}
}

\lstdefinestyle{python-style}{
    language=Python,
    style=code-common
}

\lstdefinestyle{solidity-style}{
    language=Solidity,
    style=code-common
}

\usepackage{float}

\AtBeginDocument{%
  }

\begin{document}

\title{USCSA: Evolution-Aware Security Analysis for Proxy-Based Upgradeable Smart Contracts}


\author{Xiaoqi Li}
\email{csxqli@ieee.org}
\affiliation{%
  \institution{Hainan University}
  \city{Haikou}
  \country{China}
}

\author{Lei Xie}
\email{xielei@hainanu.edu.cn}
\affiliation{%
  \institution{Hainan University}
  \city{Haikou}
  \country{China}
}

\author{Wenkai Li}
\email{cswkli@hainanu.edu.cn}
\affiliation{%
  \institution{Hainan University}
  \city{Haikou}
  \country{China}
}

\author{Zongwei Li}
\email{lizw1017@hainanu.edu.cn}
\affiliation{%
  \institution{Hainan University}
  \city{Haikou}
  \country{China}
}

\begin{abstract}
 In the case of upgrading smart contracts on blockchain systems, it is essential to consider the continuity of upgrades and subsequent maintenance. In practice, upgrade operations often introduce new vulnerabilities. Existing static analysis tools usually only scan a single version and are unable to capture the correlation between code changes and emerging risks. To address this, we propose an Upgradeable Smart Contract Security Analyzer, USCSA, which uses Abstract Syntax Tree (AST) difference analysis to assess risks associated with the upgrade process and utilizes large language models (LLMs) for assisted reasoning to achieve high-confidence vulnerability attribution.  We collected and analyzed 3,546 cases of vulnerabilities in upgradeable contracts, covering common vulnerability categories such as reentrancy, access control flaws, and integer overflow. Experimental results show that USCSA achieves a precision of 92.26\%, a recall of 89.67\%, and an F1-score of 90.95\% in detecting upgrade-induced vulnerabilities. 
As a result, USCSA provides a significant advantage to improve the security and integrity of upgradeable smart contracts, offering a novel and efficient solution for security auditing on blockchain applications.

\end{abstract}



\maketitle

\section{INTRODUCTION}
Blockchain technology continues to evolve\cite{arulprakash2022commit}, and the role of smart contracts is becoming pivotal to the digital economy\cite{zhang2025security}. The immutability inherent to smart contracts improves their transparency and trustworthiness\cite{huang2025comparative}, but contract maintenance and logic upgrades present significant challenges\cite{tonko2024visualizing}. Consequently,the proxy pattern has become a prevalent solution for achieving upgradable smart contract\cite{fang2023beyond}, as it enables logic upgrades\cite{qian2022smart} while preserving a persistent contract address\cite{mihoub2022denial}.Although this design pattern is crucial for practical deployment\cite{peng2025mining}, it also introduces new security vulnerabilities\cite{wu2025security}. If exploited, these vulnerabilities can lead to substantial financial losses\cite{uddin2024denial}.\par

Currently, Smart contract security is of critical importance\cite{yang2022multiple}.
Static analysis tools\cite{chu2023survey} and machine learning methods \cite{kiani2024ethereum}are widely employed for vulnerability detection. However,conventional static analysis tools \cite{wang2023evaluation} struggle to capture the semantic\cite{kushwaha2022systematic}  differences introduced during version iterations accurately\cite{ding2025comprehensive}. Although machine learning methods\cite{liu2025empirical} have improved the detection performance, the causal relationship problems in contract upgrades are still difficult to solve\cite{li2023novel}. Based on existing research\cite{peng2025mining,huang2024empirical,yu2025smart,shang2025cegt}, the core challenge \cite{li2023review}in smart contract security auditing is how to accurately identify security risks during version updates and clearly explain where the risks come from\cite{li2025beyond,hajihosseinkhani2025unveiling} \par

The USCSA framework directly addresses this challenge.  It employs AST differential analysis to compare structural differences across contract versions, and maps  differences to specific locations in the source code.   After static vulnerability identification, it matches the detected flaws with AST differences to determine if the vulnerabilities originate from the AST changes, thereby establishing a high-fidelity correlation between changes and potential risks.   The USCSA framework is illustrated in Figure 1. The main contributions of this paper are as follows:

\begin{itemize}
\item  We propose USCSA, which captures structural differences between versions via AST difference analysis, establishing a mapping relationship between code changes and vulnerabilities.
\item We design a semantic matching algorithm that combines multi-dimensional factors to achieve high-confidence matching between code changes and vulnerabilities.
\item USCSA integrates LLMs to refine matching results, conduct attribution analysis, and summarize patterns, establishing a rigorous security audit mechanism.

\end{itemize}
\begin{figure*}[t] 
 \centering
  \includegraphics[width=\textwidth]{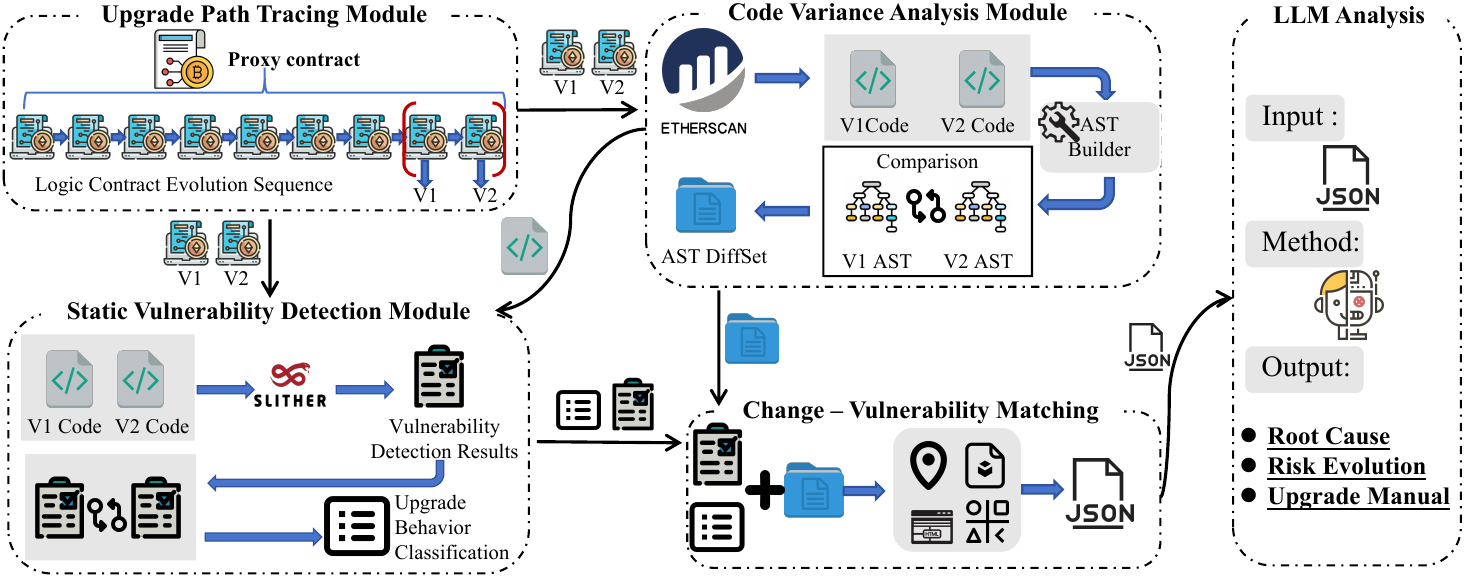} 
  \caption{USCSA Overall Framework Diagram.}
  \label{fig2}
\end{figure*}

\section{BACKGROUNDS}

\subsection{Smart Contracts and Security Audits}
Smart contracts are transparent\cite{li2025penetrating} ,automated agreements on blockchain networks\cite{shi2023automatic},they enable secure transactions in trustless environments\cite{xiang2025security} and provide economic efficiency by eliminating  intermediaries\cite{li2025facial}. With their growing adoption in finance\cite{guo2025auditing},
smart contracts have become primary targets for attacks\cite{crisostomo2025machine}. A prominent example is the 2016 DAO incident\cite{morrison2020dao}, where attackers stole Ether valued at over \$60 million by exploiting a reentrancy vulnerability. In May 2025, hackers exploited a vulnerability in the open-source library utilized by the Cetus protocol's smart contract\cite{shen2025blockchain}, resulting in losses exceeding \$220 million. These security incidents have not only resulted in financial losses\cite{huang2024revealing}, but have also eroded public trust in Blockchain\cite{zhou2022state}.

\subsection{Upgradable Smart Contracts}
Once deployed on the blockchain\cite{song2024empirical}, a smart contract is immutable\cite{gao2025implementation}.
The upgradable smart contract pattern\cite{salehi2022not} addresses this constraint by decoupling a proxy contract from a logic contract\cite{zhang2025risk}, it allows updates to the logic while maintaining a contract address. This pattern is commonly used in the OpenZeppelin \cite{amri2023review}, but there are also some risks, such as vulnerabilities arising from uninitialized or improperly reinitialized contracts\cite{ebrahimi2024large}, storage layout incompatibilities\cite{ruaro2024not}, and access control flaws of upgrading without authorization\cite{bodell2023proxy}.

\subsection{Related Work}
In  smart contract vulnerability detection,Jaeseung Choid et al. proposed the Smartian tool\cite{choi2021smartian}, which combines static and dynamic data flow analysis to improve the efficiency of smart contract fuzz testing. Asem Ghaleb et al. evaluated the effectiveness of static analysis tools \cite{ghaleb2020effective} like Oyente, Mythril, and Slither through vulnerability injection methods. Xueyan Tang et al.'s Lightning Cat framework \cite{tang2023deep} employs optimized CodeBERT, LSTM, and CNN models to detect smart contract vulnerabilities, reducing false positives and false negatives through semantic analysis. Wei Ma et al. proposed the two-stage fine-tuning framework iAudit\cite{ma2024combining}, constructing an audit framework that accurately detects smart contract vulnerabilities while generating high-quality, interpretable analyses.Eshghie et al. designed SoliDiffy, an AST differential tool specifically for Solidity, which can generate precise editing scripts\cite{eshghie2024solidiffy} .However, these efforts\cite{javaid2022review,gadekallu2022blockchain,li2022ether,sapra2023impact,gallenmuller2021pos，qi2024sok,li2022sok,taherdoost2023smart} have not fully integrated AST differential analysis with the mapping between code changes and vulnerabilities. Our research fills this gap.

\section{METHODOLOGY}

\subsection{Upgrade Path Tracing Module}
To obtain the required latest version V1 and the subsequent version V2, USCSA tracks the change history of the logic contracts of proxy contracts using event logs, and then acquires information such as the source code of V1 and V2 through Etherscan\cite{wang2022dao}. After obtaining the change history, we designed a dual sorting algorithm, combining version number semantic parsing and time series 
analysis to determine the priority of contract versions; we also built a data fusion pipeline and an utomated downgrade mechanism to ensure reliability and accuracy when analysing proxy contracts at scale.
\subsection{AST Difference Analysis Module}
After obtaining  contract data, USCSA builds an AST for differential analysis to capture structural changes between versions.We employ a dual-engine AST construction approach: the primary  engine utilizes the Falcon framework, and the backup engine relies on ANTLR4\cite{chen2021conversion}.After generation, the system evaluates the quality of the AST using indicators such as node integrity, structural integrity, and semantic integrity. This  configuration enhances the system's fault tolerance. To detect AST difference between V1 and V2, we use the GumTree algorithm to compare the tree structure . This method extracts  semantic information such as function and variable identifiers, and maps structural changes to modifications at the code level, providing  support for subsequent vulnerability analysis\cite{de2023distributed}.

\subsection{Static Vulnerability Detection Module}
Table 1 summarizes the vulnerability detection results of V1 and V2. 
USCSA mainly focuses on two key scenarios: one is that V1 has no vulnerabilities but V2 has added vulnerabilities, which can directly explain the causal relationship of upgrade risks.Another situation is that V1 has a vulnerability but V2 has fixed it, which is helpful for optimizing the security upgrade strategy. The remaining scenarios serve as control groups, providing both positive and negative data support. This data establishes a foundation for subsequent behavior attribution and risk inference using LLMs.

\begin{table}[htbp] 
\centering
\small
\setlength{\tabcolsep}{6pt}
\renewcommand{\arraystretch}{1.2}
\begin{tabular}{c c c c }
\hline
\textbf{v1 Bug} & \textbf{v2 Bug} & \textbf{Upgrade Behavior} & \textbf{Security Conclusion}\\
\hline
$\times$ & $\checkmark$ & Introduce Vulnerability & Risk Increased  \\
$\checkmark$ & $\times$ & Fix Vulnerability & Security Improved \\
$\times$ & $\times$ & Smooth Upgrade & No Security Impact  \\
$\checkmark$ & $\checkmark$ & Invalid Upgrade & Uncertain, needs analysis   \\
\hline
\end{tabular}
\caption{Upgrade Behaviors and Security Classification}
\label{tab:upgrade_behavior}
\Description{This table classifies the security impact of software upgrades based on the presence (checkmark) or absence (cross) of a specific bug in the old (v1) and new (v2) versions.}
\end{table}

\subsection{Change-Vulnerability Matching Module}
To address the lack of direct correlation analysis between code changes and generated vulnerabilities\cite{fei2023msmart}, this study proposes an enhanced semantic matching algorithm that combines multi-dimensional information to achieve high-confidence attribution of vulnerability changes. As shown in Algorithm 1. The algorithm evaluates the correlation between code changes and vulnerabilities from the following four perspectives:

\begin{itemize} 
    \item \textbf{Position Matching}: Analyzes the proximity of the physical location of the change and the vulnerability in the code. 
    \item \textbf{Pattern Matching}: Identifies the structure of the change statement using the vulnerability pattern library. 
    \item \textbf{Semantic Matching}: Measures semantic associations from dimensions such as function name, node type, keyword overlap, variable names, change type , common vulnerability characteristics, and impact areas.
    \item \textbf{Change Type Matching}: Considering the correspondence between the nature of change operations in the AST and vulnerability types.
\end{itemize} 

For any change $c_i$ and vulnerability $v_j$, the confidence level of their matching is calculated as a weighted sum,the weights are empirically assigned. where $S_*$ represents the matching score of one dimension and has been normalized to the range of $[0,1]$:
\begin{small}
\begin{equation} \label{eq:confidence}
C(c_i, v_j) = \min\Bigl(1.0, \; 0.3 \cdot S_{\text{pos}} + 0.25 \cdot S_{\text{pattern}} + 0.3 \cdot S_{\text{semantic}} + 0.15 \cdot S_{\text{type}}\Bigr)
\end{equation}
\end{small}
The dimension scores are defined as follows:
\begin{itemize}
    \item \textbf{Position Score}: Let $linedist = |loc(c_i) - loc(v_j)|$ be the absolute line number difference.
    \begin{equation} \label{eq:position_score} 
     S_{\text{pos}} = 
     \begin{cases} 
     1.0, & \text{if } linedist = 0 \\ 
     0.8, & \text{if } 0 < linedist \leq 2 \\ 
     0.5, & \text{if } 2 < linedist \leq 5 \\ 
     0.2, & \text{if } 5 < linedist \leq 10 \\ 
     0.1, & \text{otherwise.} 
     \end{cases} 
    \end{equation}
    
    \item \textbf{Pattern Score}: Based on keyword overlap in a $\pm 5$-line context window around $c_i$.
    \begin{equation} \label{eq:pattern_score}
    S_{\text{pattern}} = \min(1.0, \; \text{count}_{\text{common\_keywords}} \times 0.1)
    \end{equation}
    
    \item \textbf{Semantic Score}: A composite of six weighted features:
    \begin{small}
    \begin{equation} \label{eq:semantic_score} 
    S_{\text{semantic}} = \min(1.0, \; 0.3 F_1 + 0.2 F_2 + 0.15 F_3 + 0.15 F_4 + 0.1 F_5 + 0.1 F_6) 
    \end{equation}
    \end{small}
    \begin{itemize}
        \item $F_1$: Fuzzy function name matching.
        \item $F_2$: AST node type relevance.
        \item $F_3$: Keyword overlap in descriptions.
        \item $F_4$: Change operation type similarity.
        \item $F_5$: Common vulnerability trait matching.
        \item $F_6$: Change impact area.
    \end{itemize}

    \item \textbf{Change Type Score ($S_{\text{type}}$)}: Derived from a predefined mapping between AST change operation types and vulnerability types, with scores in $[0,1]$.
\end{itemize}

\begin{algorithm}[htbp]
\small
\caption{Enhanced Vulnerability Matching Algorithm}
\label{alg:matching}
\SetAlgoLined
\KwIn{AST changes $ast\_changes$; Vulnerabilities $vulnerabilities$; Source code $source\_code$}
\KwOut{Matched pairs $matched\_pairs$}
\BlankLine

\ForEach{change $c_i$ in $ast\_changes$}{
  \ForEach{vulnerability $v_j$ in $vulnerabilities$}{
    $S_{\text{pos}} \gets \text{CalcPos}(c_i, v_j)$
    $S_{\text{pattern}} \gets \text{CalcPattern}(c_i, v_j, source\_code)$ 
    $S_{\text{semantic}} \gets \text{CalcSemantic}(c_i, v_j)$ 
    $S_{\text{type}} \gets \text{CalcType}(c_i, v_j)$ 
    
    \BlankLine
    \tcp{Compute matching confidence}
    $C(c_i, v_j) \gets \min\Bigl(1.0, \; 0.3 \cdot S_{\text{pos}} + 0.25 \cdot S_{\text{pattern}} + 0.3 \cdot S_{\text{semantic}} + 0.15 \cdot S_{\text{type}}\Bigr)$
    \BlankLine
    
    \If{$C(c_i, v_j) > 0.6$}{
      Add $(c_i, v_j, C(c_i, v_j))$ to $matched\_pairs$\;
    }
  }
}
\Return{sorted($matched\_pairs$, by=confidence, descending=True)}
\end{algorithm}

\subsection{LLM-assisted “Change-Vulnerability” Causality Matching}
We use the code understanding and reasoning capabilities of  LLM to analyze the matching results.By designing structured prompts, LLM is guided to analyze the vulnerabilities in five dimensions:
 \begin{itemize}
    \item   Root cause analysis of vulnerabilities;
    \item   Assessment of the impact of code changes on security;
    \item   Correlation analysis of change types and vulnerability types;
    \item   Automatic assignment of risk pattern labels;
    \item   Targeted remediation recommendation generation.

\end{itemize}
The LLM's analysis results are output in JSON format,  facilitates subsequent data processing and pattern recognition.By comparing and analyzing the changes in the distribution of vulnerabilities in the versions before and after the upgrade, we establish a risk evolution pattern recognition mechanism,that automatically recognizes three types of risk changes: Introduce Vulnerability, Fix Vulnerability, and Invalid Upgrade. Based on these patterns,system is able to generate risk evolution analysis and escalation  assessment reports\cite{kushwaha2022ethereum}.
We designed a upgrade bug pattern extraction algorithm based on LLM.By analyzing escalation cases, LLM identifies common escalation error patterns, such as missing privilege control\cite{perez2022secure}, unsynchronized state variables\cite{krichen2022formal}, and external call risks.These patterns are standardized as risk labels, forming an upgrade security checklist.Based on the accumulation of  analysis results, the system automatically generates an upgrade recommendation manual. The prompt template is shown in Figure 2.

\begin{figure}[h]
\centering
\includegraphics[width=8cm]{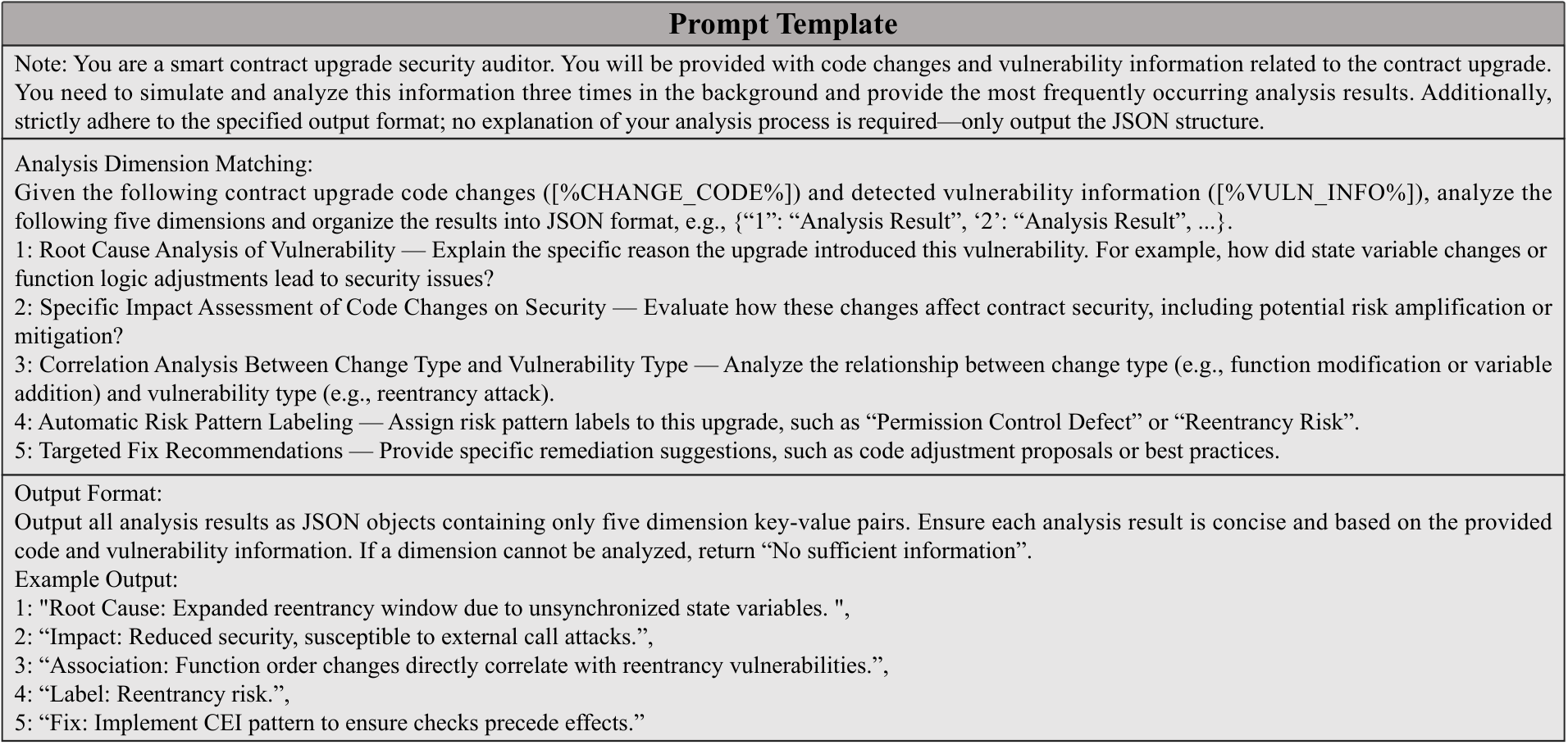}
\caption{LLM-Assisted Analysis Prompt.}
\label{fig3}
\end{figure}
\section{RESULT}
\subsection{Dataset}
We collected and analyzed 3,546 upgradable smart contract cases.From these cases, we extracted the core features used for vulnerability matching and upgrade classification, including the scale of code modifications, the distribution of vulnerability types, the upgrade patterns, the proximity of locations, the semantic similarity, and the confidence score between modifications and vulnerabilities.These features reveal the patterns of vulnerability introduction and remediation.We obtained the source code, transaction logs, and upgrade events via the Etherscan API. By combining the Slither with AST differential analysis, we parsed the ASTs and standardized the relevant features to ensure data consistency and reusability.
\subsection{Performance and Evaluation}
USCSA analyzed 3,546 proxy contracts and saved the detailed information of the successfully analyzed contracts as JSON reports for subsequent optimization. 
Enabling an enhanced matching mode allows the system to identify additional semantic-level vulnerabilities, increasing detection coverage by approximately 15\%. For vulnerability detection, USCSA combined the Slither with an enhanced semantic matcher to discover several types of vulnerability specific to upgradeable contracts, including proxy storage collisions, newly introduced upgrade vulnerabilities, and behavioral inconsistencies.\par

The experimental results show that among the 3,546 analyzed contracts, a total of 4,872 vulnerability instances were identified, with an average of 1.37 vulnerabilities per contract. Among them, the proportions of high-risk, medium-risk and low-risk vulnerabilities are approximately 28\%, 52\% and 20\% respectively. The results are shown in Figure 3. \par
\begin{figure}[htbp]
\centering
\includegraphics[width=8cm]{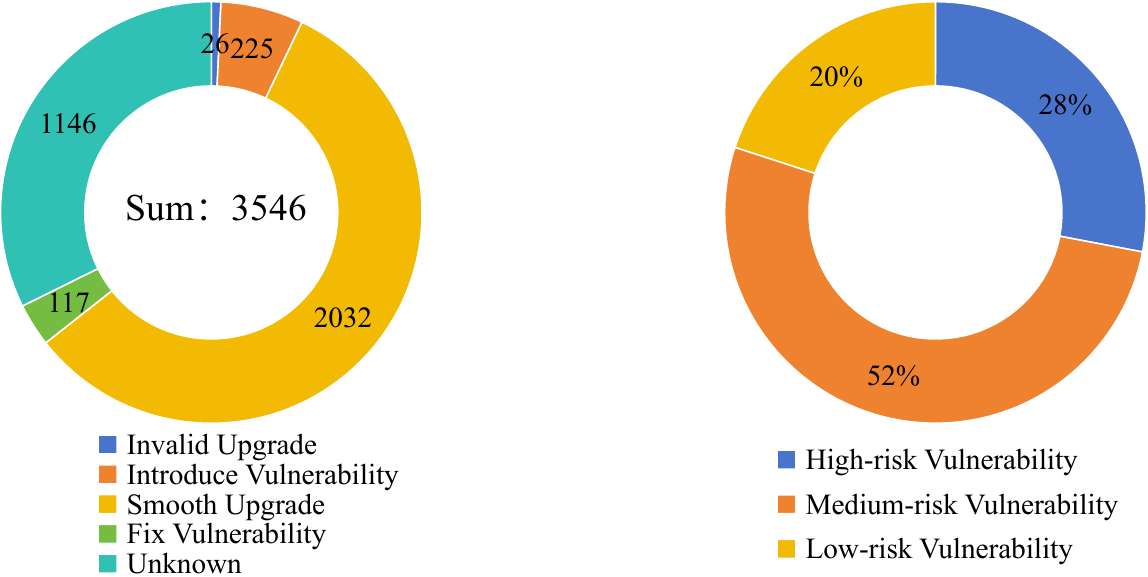}
\caption{USCSA Test results.}
\label{fig1}
\end{figure}
We manually verified 500 randomly sampled contracts: results show 92.26\% precision, 89.67\% recall, and 90.95\% F1-score (Table 2). Compared with conventional tools like Slither, USCSA  Matching module improves detection accuracy for upgrade-induced vulnerabilities by approximately 18\% and shows superior performance in handling multi-version upgrade paths.
\begin{table}[htbp]
\centering
\caption{Evaluation Results of USCSA}
\label{tab:uscsa-results}
\begin{tabular}{l c}
\hline
\textbf{Metric} & \textbf{Value} \\
\hline
Precision & 92.26\% \\
Recall    & 89.67\% \\
F1-score  & 90.95\% \\
\hline
\end{tabular}
\Description{Performance metrics of the USCSA framework}
\end{table}

\section{Discussion and Future Plans}
The USCSA proposed in this study establishes a mapping  between modified code and vulnerabilities by integrating AST differential analysis with change-vulnerability matching.  It subsequently achieves high-confidence matching  through LLM-assisted attribution analysis.
Experiments show that USCSA demonstrates excellence in  detection accuracy and change-matching efficiency, and has research and practical application value. It also provides a reproducible and quantifiable upgrade security audit process.\par
The USCSA also has some limitations: it is currently mainly focused on Ethereum, and its capabilities in different scenarios need further exploration; during attribution analysis, instability exists in complex contract scenarios, requiring more data support.\par
Future directions include expanding USCSA to other blockchain platforms such as Solana and Polkadot to support cross-chain and cross-version upgrade analysis; optimising semantic matching algorithms to improve the confidence of code change and vulnerability correlation; extending data cases to the DeFi and NFT sectors, collecting diverse upgrade examples to build a knowledge base and upgrade guidelines, further reducing the cost of manual auditing.

\bibliographystyle{ACM-Reference-Format}
\bibliography{main}

\appendix

\end{document}